\documentclass[12pt,preprint]{emulateapj}
\usepackage{amsmath, natbib, placeins}
\begin{document}


\title{Probing the Pulsar Origin of the Anomalous Positron Fraction \\ with AMS-02 and Atmospheric Cherenkov Telescopes}
\author{Tim Linden$^{1}$ and Stefano Profumo$^{1,2}$}
\affil{$^1$ Department of Physics, University of California, Santa Cruz, 1156 High Street, Santa Cruz, CA, 95064}
\affil{$^2$ Santa Cruz Institute for Particle Physics, University of California, Santa Cruz, 1156 High Street, Santa Cruz, CA, 95064}
\shortauthors{}

\begin{abstract}
Recent observations by PAMELA, Fermi-LAT, and AMS-02 have conclusively indicated a rise in the cosmic-ray positron fraction above 10 GeV, a feature which is impossible to mimic under the paradigm of secondary positron production with self-consistent Galactic cosmic-ray propagation models. A leading explanation for the positron fraction rise is an additional source of electron-positron pairs, for example one or more mature, energetic, and relatively nearby pulsars. We point out that any one of two well-known nearby pulsars, Geminga and Monogem, can satisfactorily provide enough positrons to reproduce AMS-02 observations. A smoking-gun signature of this scenario is an anisotropy in the arrival direction of the cosmic-ray electrons and positrons, which may be detectable by existing, or future, telescopes. The predicted anisotropy level is, at present, consistent with limits from Fermi-LAT and AMS-02. We argue that the large collecting area of Atmospheric Cherenkov Telescopes (ACTs) makes them optimal tools for detecting such an anisotropy. Specifically, we show that much of the proton and $\gamma$-ray background which affects measurements of the cosmic-ray electron-positron spectrum with ACTs may be controlled in the search for anisotropies. We conclude that observations using archival ACT data could already constrain or substantiate the pulsar origin of the positron anomaly, while upcoming instruments (such as the Cherenkov Telescope Array) will provide strong constraints on the source of the rising positron fraction.
\end{abstract}

\maketitle

\section{Introduction}

Local observations of cosmic-ray fluxes are among the most powerful probes of high-energy astrophysical processes in our Galaxy. These measurements can probe both the nature of particle production in energetic astrophysical systems, as well as the physics of particle diffusion through the Galaxy. Recently, results from the Advanced Thin Ionization Calorimeter (ATIC) balloon experiment~\citep{atic_electron_bump}, as well as from the Payload for Antimatter Matter Exploration and Light-nuclei Astrophysics (PAMELA) satellite~\citep{pamela_positron_excess}, indicated possible anomalies in the expected flux of cosmic ray leptons. Specifically, ATIC observed a bump in the cosmic-ray electron-plus-positron ($e^\pm$) spectrum at an energy of approximately 800~GeV, the intensity of which has since been cast into doubt by observations with the High Energy Stereoscopic System (H.E.S.S.) telescope~\citep{hess_low_energy} and the Fermi Large Area Telescope (Fermi-LAT)~\citep{fermi_electron_spectrum}. PAMELA, on the other hand, using a built-in magnetic field that allows it to differentiate the charge of cosmic-ray species, observed a rise in the positron fraction of the cosmic-ray $e^\pm$ spectrum above 10~GeV. This result has been confirmed by subsequent observations using the Fermi-LAT, cleverly employing the geomagnetic field structure and the Earth shadow to distinguish cosmic-ray particle charges~\citep{fermi_positron_fraction}.

Very recently, the Alpha Magnetic Spectrometer (AMS-02) released its initial findings~\citep{ams_data}, which confirm the previously observed rising positron fraction with significantly better statistical and systematic errors than previous experiments. Additionally, AMS-02 extended the energy range of the positron fraction up to an energy of 350~GeV, finding that while the positron fraction continued to rise at these energies, the slope of the positron fraction appears to flatten above $\sim$150~GeV. This suggests that the positron fraction may either approach a stable asymptotic value, or that it might even decrease at higher energies. However, no evidence of a sharp cutoff in the positron fraction was observed. Compared with PAMELA measurements, AMS-02 results also find an anisotropy which is approximately 20\% smaller at high energy. Since the secondary positron signal in this range is extremely small, this results in an approximately 20\% reduction in the primary positron flux necessary to explain the AMS-02 positron ratio in light of Fermi-LAT constraints on the total electron plus positron spectrum compared to fits to the PAMELA measurements. In addition, the slightly softer spectrum inferred from the AMS-02 data compared to the PAMELA measurement also points to a smaller normalization of the needed additional primary component.

Myriad models have been formulated in order to explain the rise in the positron fraction, nearly all of which require a new injection spectrum of primary positrons~\citep{2011APh....34..528D}. These models can be broken down into two major categories. The majority (too numerous to list here) concern the annihilation of dark matter particles into, primarily if not exclusively, leptonic final states (e.g. \citet{a_theory_of_dark_matter, cholis_pamela, cirelli_pamela}). Relevant dark matter models almost uniformly require a massive dark matter particle (M$_\chi~\gtrsim$~0.5~TeV, possibly even larger after the AMS-02 data extended the positron fraction to 350 GeV) with an annihilation cross-section that significantly exceeds the thermal annihilation cross-section implied by the ``WIMP miracle''. Since leptophilic dark matter annihilation should additionally create a bright $\gamma$-ray signal (for example from inverse Compton up-scattering of background radiation fields, or from final state bremsstrahlung, or from the hadronic decays of $\tau$ leptons), many of these models have been ruled out by Fermi-LAT $\gamma$-ray observations of, e.g., the Galactic center, dwarf spheroidal galaxies or nearby galaxy clusters~\citep[see e.g.][for early studies]{pamela_dark_matter_gamma_constraints, hooper_linden, hooper_gc_constraints, fermi_lat_dwarf_spheroidals, Ackermann:2010rg}. Other constraints arise from distortions to the CMB spectrum or from radio observations. These analyses, will need to be updated in view of the new AMS-02 data, which surely restrict the class of dark matter models which can explain the positron fraction solely based on the observed positron fraction spectral shape. However, this task lies beyond the scope of the present study.

A second class of models relies on mature (i.e. in an age range between $10^4$ and $10^6$ yrs \citep{Grasso:2009ma}) pulsars, which are known sources of $e^\pm$ pairs, in order to produce the rising positron fraction. Models of particle creation in pulsars' magnetospheres indicate that these objects produce a significant population of energetic leptons with a hard spectrum~\citep{ hooper_pulsars_pamela, delahaye_local_pulsars, profumo_occams_razor, malyshev_local_pulsars, barger_pulsars_pamela, 2009arXiv0912.0264K}. The pulsar contribution to the positron fraction could stem from the combined fluxes of a Galactic population of pulsars~\citep[see e.g.][]{barger_pulsars_pamela}, or may be dominated by a single, nearby source~\citep[e.g.][]{profumo_occams_razor, 2009PhRvL.103e1101Y}.

Pinpointing the origin of the rising positron fraction requires a method of differentiating pulsar and dark matter models for this signal. While each class of models may produce slightly different predictions for high energy diffuse radiation (e.g. from inverse Compton emission), our poor knowledge of the diffuse $\gamma$-ray sky has precluded this avenue from conclusively distinguishing between models. Similarly, accurate measurements of the total $e^\pm$ spectrum can hardly produce a smoking gun favoring either model: in the dark matter case, a spectral cutoff in the positron fraction emerges at the kinematic edge corresponding to the dark matter particle mass; in the case of pulsars, a cutoff is predicted from energy losses and diffusion and/or from the intrinsic source injection spectrum. An observational test that could conclusively rule out a dark matter interpretation does, however, exist; the detection of an anisotropy in the arrival direction of cosmic-ray leptons.

Models using dark matter to explain the cosmic-ray excess are unlikely to produce significant anisotropies in the arrival direction of the $e^\pm$ observed at the solar position. A dark matter clump that produced a large enough anisotropy would also produce a bright gamma-ray emission that would be easily and distinctively detectable by the Fermi-LAT \citep{Brun:2009aj}. The anisotropy induced by nearby pulsars may instead be significant and potentially observable \citep{hooper_pulsars_pamela}, although the anisotropy may be highly suppressed if numerous pulsars located randomly in the sky contributed significantly to the overall positron flux. \citet{profumo_occams_razor}, however, pointed out that the expected anisotropies from selected nearby pulsars (e.g. Geminga, Monogem, Vela, CTA1, B0355+54) may be detectable at a 2$\sigma$ level with 5 years of Fermi observations. A search using one year of Fermi-LAT data did not uncover any signs of anisotropy~\citep{fermi_1yr_electron_anisotropy}, consistent with expectations from pulsar models. In fact, the current constraints lie approximately an order of magnitude above pulsar projections.


Here, we ague that Atmospheric Cherenkov Telescopes (ACTs) offer a very promising tool in searches for cosmic-ray $e^\pm$ anisotropies associated with a local source. Due to their large effective area, ACTs are capable of detecting much smaller anisotropies than space-based telescopes. However, ACTs also face significant hurdles due to the systematics associated with differentiating $e^\pm$ showers from hadronic showers created by high energy protons, which dominate the $e^\pm$ flux by several orders of magnitude: Uncertainties in hadronic shower rejection present a systematic error of approximately 30\% in the cosmic-ray $e^\pm$ flux observed by H.E.S.S., an uncertainty which is several times larger than those from the Fermi-LAT cosmic-ray $e^\pm$ measurement at similar energies.

In this \emph{paper}, we first point out that a simple model employing a single nearby pulsar in addition to the best-fit diffuse cosmic-ray background produces an extraordinarily good fit to the positron fraction observed by AMS-02 and is consistent with the  combined electron plus positron flux as measured by the Fermi-LAT and by H.E.S.S.  The AMS-02 data are perfectly compatible with estimates of the $e^\pm$ flux from nearby mature pulsars such as Geminga and Monogem, although other pulsars may also contribute. We  note that these pulsars are expected to produce a significant anisotropy in the cosmic-ray $e^\pm$ spectrum, and evaluate the associated signal. While this anisotropy falls nearly an order of magnitude below the current constraints from both AMS-02 and the Fermi-LAT, we point out that the large effective area of ACTs may produce significantly more robust constraints. Specifically, we show that as long as the hadronic background observed by these telescopes is highly isotropic, it can be treated as a statistical, rather than a systematic uncertainty. Already existing ACT observations are likely to significantly cut into the scenarios where, for example, the Monogem pulsar is responsible for the rising positron fraction. Finally, we note that the upcoming Cherenkov Telescope Array (CTA) will greatly enhance the potential for searches for cosmic-ray $e^\pm$ anisotropies, due to both its improved effective area and hadronic rejection efficiency.

The outline of this paper is as follows:  In Section~\ref{sec:flux}, we discuss the contributions of young, known pulsars to the total electron plus positron flux observed at the solar position. In Section~\ref{sec:acts} we compute the anisotropy resulting from the Geminga and Monogem pulsars which we find to provide the best fits to the observed cosmic-ray signal, and we discuss the possibility for detecting this anisotropy with ACTs.  In Section~\ref{sec:results} we present our results on anisotropy searches with ACTs, and discuss possible systematics resulting from Galactic diffuse emission in Section~\ref{sec:diffuse}. Finally, in Section~\ref{sec:discussions} we discuss the implications of our results and conclude.

\section{The Contribution of Nearby Pulsars to the Diffuse Electron-Positron Flux}
\label{sec:flux}

In order to calculate the background cosmic-ray electron and positron flux attributable to diffuse Galactic processes, we employ the \emph{Galprop} cosmic-ray propagation code~\citep{galprop, galprop_recent}. \emph{Galprop} self-consistently calculates the injection and diffusion of cosmic-ray electrons, positrons, protons, and nuclei throughout the interstellar medium of the Milky Way galaxy, taking into account effects such as convection and re-acceleration, as well as energy loss processes such as synchrotron and bremsstrahlung emission. The \emph{Galprop} code has been specifically tailored to accurately reproduce the ratios of radioactive cosmic-ray primary and secondary nuclei (e.g. the Boron to Carbon ratio) which currently stand as the best handles on the parameters of cosmic-ray propagation in the Milky Way. Specifically, we employ here the best-fitting Bayesian model of \citet{trotta_bayesian}, which finds a posterior mean for the diffusion constant of 8.32$\times$10$^{28}$~cm$^2$s$^{-1}$, normalized at 4 GeV, with an energy index $\delta$~=~0.3, nearly equivalent to that first proposed by~\citet{kolmogorov}. We employ all parameters from this model in the following analysis, but linearly vary the total injection normalization of cosmic-rays in order to create a best fitting model when various pulsar profiles are added to account for additional sources of electrons and positrons\footnote{In practice, we adopt a simple rescaling of the \emph{Galprop} diffuse $e^\pm$ flux by a factor 0.8.}


\begin{figure*}
                \plottwo{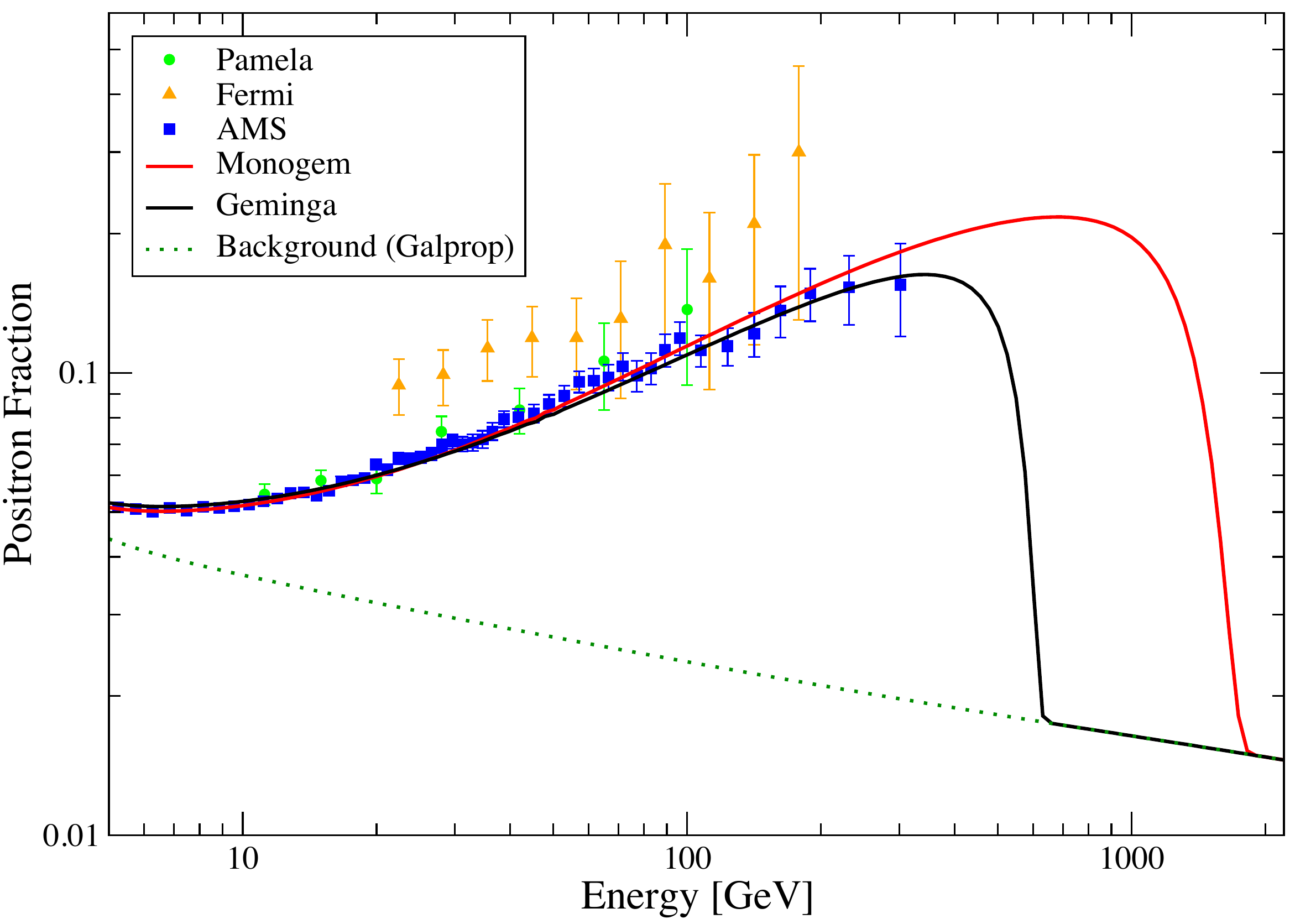}{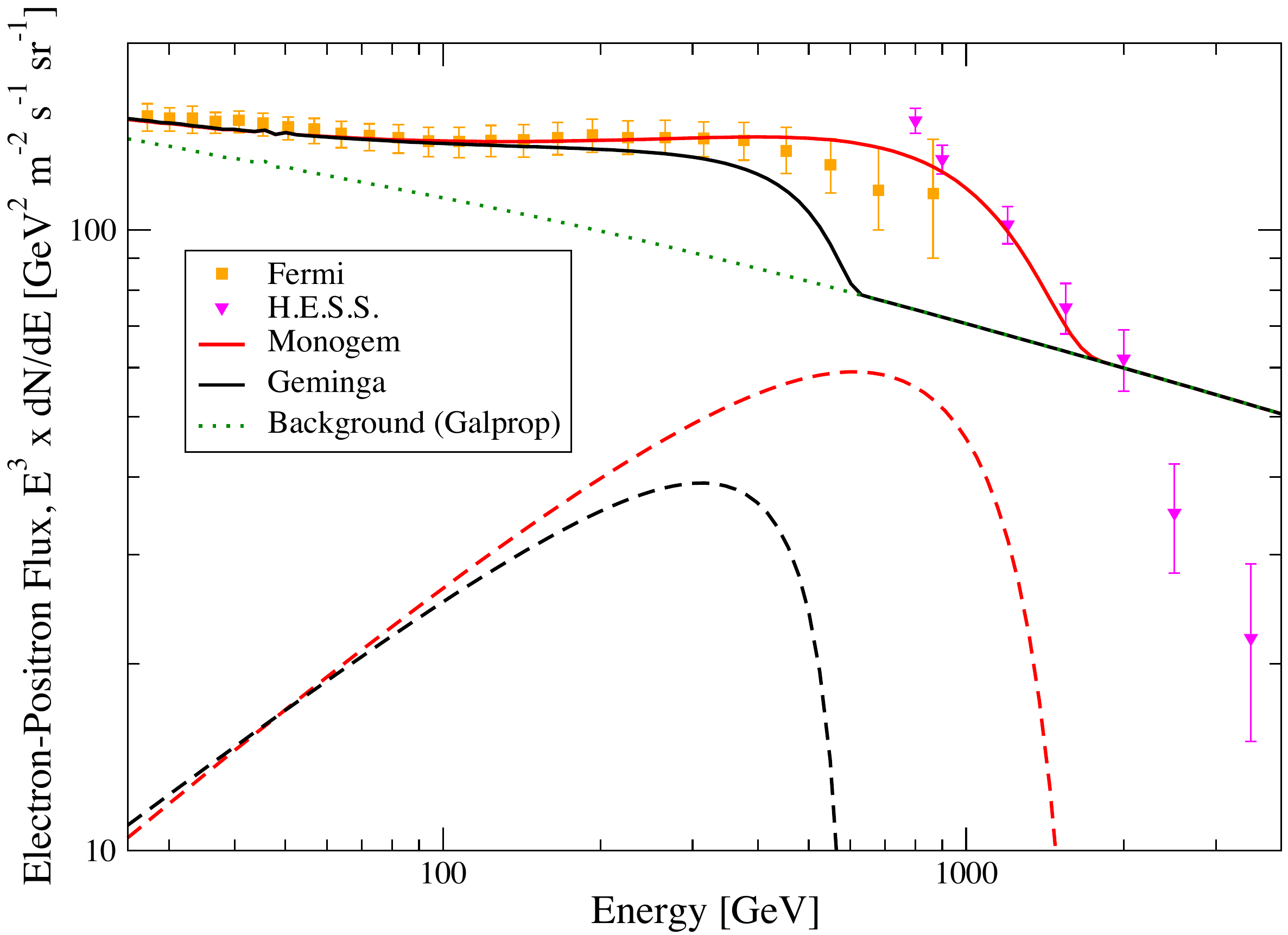}
                	\caption{ \label{fig:flux} Left: The positron fraction from a combination of the {\em Galprop} model for the diffuse $e^\pm$ Galactic background (green dotted), along with contributions from the Geminga (black) and Monogem (red) pulsars, compared with data from PAMELA (green circles), Fermi-LAT (orange triangles) and AMS-02 (blue squares). Right: The flux of cosmic-ray electrons and positrons from a combination of the same \emph{Galprop} model (green dotted), with contributions from the Geminga (black dashed) and Monogem (red dashed) pulsars. These create a total cosmic-ray lepton spectrum (black and red solid respectively), which can be compared with data from the Fermi-LAT (orange squares) and H.E.S.S. (pink diamond) observations, (right). Note that the diffuse background from \emph{Galprop} was not tuned to reproduced the H.E.S.S. data, and we do not attempt to fit those data above 1 TeV.}
\end{figure*}

We calculate the contribution to high-energy cosmic-ray electrons and positrons from pulsars in the context of the point-like and burst-like approximation of \citet{1970ApJ...162L.181S} (see also \citet{Panov:2013fma} for a recent review). We assume that pulsars inject in the interstellar medium (ISM) as many electrons as positrons, with a source function
\begin{equation}\label{eq:source}
Q(\vec r,t,E)=Q(E))\delta^3(\vec r)\delta(t)),
\end{equation}
with an exponentially suppressed power-law spectrum parameterized by
\begin{equation}
Q(E)=Q_0E^{-\gamma}\exp\left(E/E_{\rm cut}\right).
\end{equation}
The normalization $Q_0$ of the source function is then associated with the total energy output of the pulsar via the simple relation
\begin{equation}
\int_0^\infty\ Q(E)E{\rm d}E=\eta W_0,
\end{equation}
with $W_0$ indicating the total pulsar energy output, and $\eta<1$ the relative efficiency with which such energy is converted into the energy associated with $e^\pm$ pairs injected in the ISM.

The source function of Eq.~(\ref{eq:source}) is substituted in the relevant differential equation describing cosmic-ray transport in the limit where the energy loss terms and the diffusion constant are both independent of position, yielding a simple, exact analytic solution \citep[see e.g.][]{1995PhRvD..52.3265A}. Among the remarkable features of this asymptotic solution is that there exists a cut-off in energy set by the $e^\pm$ inverse Compton and Synchrotron energy losses and by the time $T$ of the burst-like injection. If the energy loss term is cast as
\begin{equation}
\frac{{\rm d}E}{{\rm d}t}=-b_0\ E^2,
\end{equation}
the cutoff corresponds to $E_{\rm loss}\simeq1/(b_0 T)$. Here, we employ $b_0=1.4\times 10^{-16}\ {\rm GeV}^{-1} {\rm s}^{-1}$. This implies $E_{\rm loss}\simeq 750$ GeV for $T=300$ kyr, and $E_{\rm loss}\simeq 226$ GeV for $T=1$ Myr. On the other hand, there exists a second exponential suppression, due to $e^\pm$ diffusion during the relevant energy-loss time-scale, of the form
\begin{equation}
\exp\left(-\left(d/r_{\rm diff}(E)\right)^2\right),
\end{equation}
where $d$ is the distance to the (point-like) source and $r_{\rm diff}(E)$ is a characteristic, energy-dependent diffusion length. For example, at energies around 1 TeV the typical diffusion length falls well within 1~kpc. These two constraints effectively limit the class of pulsars that could add to the positrons feeding the anomaly observed in the rising positron fraction. In particular, the age of the candidate pulsar must fall between the time-scale for the confinement of  $e^\pm$ pairs within the nebula ($\sim$~$10^4$ yr), and the time-scale corresponding to energies large enough to explain the highest energy where the positron anomaly is observed ($\sim$~1~Myr for $E\sim350$~GeV). Pulsars in this age range are usually dubbed ``mature'' \citep{Grasso:2009ma}.

While we do not attempt here to exhaustively provide a census of the known pulsars that could contribute to the positron fraction anomaly, this exercise has been carried out in the past for the PAMELA data (e.g. \citet{profumo_occams_razor}, \citet{Grasso:2009ma}, \citet{malyshev_local_pulsars}, \citet{hooper_pulsars_pamela} among others). Instead, we provide two examples of well-known pulsars in the right distance and age range to potentially contribute very significantly to (or as we shall see, explain completely!) the recent high-statistics AMS-02 data \citep{ams_data}. 

We focus here on two of the most luminous nearby pulsars in the correct age and distance range: the Geminga (J0633+1746) pulsar and the Monogem (B0656+14) pulsar\footnote{Other radio (or $\gamma$-ray only) pulsars are also capable of contributing significantly to the local positron flux. A more thorough discussion of such candidates can be found in \citet{Gendelev:2010fd}}. We assume that the burst-like injection happened for both pulsars at look-back times comparable to the age of the pulsar (i.e. we neglect the trapping time of the $e^\pm$ in the nebula compared to the age of the pulsar, see e.g. \citet{Grasso:2009ma}). We set Geminga's age to $3.42\times 10^5$ yrs, and Monogem's to $1.11\times 10^5$ yrs, and the distances to each pulsar to 0.15 and 0.29 kpc, respectively \citep[see][and references therein]{profumo_occams_razor}. 

While the age and distance for the pulsars is set by observation, less is known about the spectral shape and energy output. The exponential cutoff $E_{\rm cut}$ is entirely immaterial for Geminga, as the corresponding $E_{\rm loss}\ll E_{\rm cut}$, while for the purpose of reproducing the AMS-02 results $E_{\rm cut}$ is also irrelevant for Monogem, unless it falls in the sub-TeV domain, which is theoretically implausible \citep{malyshev_local_pulsars}. For definiteness, we set for both pulsars $E_{\rm cut}=2$ TeV (changing this parameter clearly affects the details of the spectrum of $e^\pm$ at $E\gtrsim E_{\rm cut}$, but reproducing such spectrum in detail is not within the scopes of this study, as it is largely affected by the diffuse Galactic background, not accurately modeled by \emph{Galprop} at these energies).

Direct observations of the pulsar source spectral index $\gamma$ are available for a handful of pulsar wind nebul\ae~\citep[see e.g.][]{Green:2009qf}, and point to $\gamma<2$, although in some cases $\gamma\sim2$ is also observed \citep{Aharonian:2006xx}. We find that, for our choices of diffusion parameters, the AMS-02 data are best fit by $1.8\lesssim \gamma\lesssim 2$, and we employ $\gamma=1.9$ for Geminga and $\gamma=1.95$ for Monogem. The resulting normalizations required to provide a fit to the AMS-02 data with a single pulsar correspond to $\eta W_0=2\times 10^{49}$ erg for Geminga and to $\eta W_0=8.6\times 10^{48}$ erg for Monogem. Within the context of our diffusion model, we note that these values act as upper limits on the total lepton flux from each pulsar for any scenario which is compatible with the AMS-02 data, since these values must decrease if additional sources are considered. The total energy outputs we find depend quite sensitively on the assumptions made for the spectral slope, but are generically compatible with the total energy output expected from a mature pulsar, which ranges within $5\times 10^{48}\lesssim W_0/{\rm erg}\lesssim 5\times 10^{50}$, \citep{Delahaye:2010ji, malyshev_local_pulsars}.

Employing a combination of the  \emph{Galprop} Galactic $e^\pm$ diffuse background model, rescaled by a factor 0.8 to account for the additional sources, and the calculated flux from each candidate pulsar, in Figure~\ref{fig:flux} we show the positron fraction (left) and the combined flux of electrons and positrons (right) observed at the solar position for models in which  the Geminga pulsar dominates the production of nearby positrons (black), and a model where the Monogem pulsar dominates cosmic-ray positron production (red). In each case, we find an extremely good match between our results and AMS-02 observations. 


\section{Detection of a Cosmic-Ray Electron/Positron Anisotropy with ACTs}
\label{sec:acts}

In the context of diffusive propagation, we estimate the expected anisotropy from a source at a distance $d$ that injected $e^\pm$ at a time $T$ \citep[e.g.][]{Grasso:2009ma} with
\begin{equation}
\Delta=\frac{3}{2c}\frac{d}{T}\frac{(1-\delta)E/E_{\rm loss}}{1-\left(1-E/E_{\rm loss}\right)^{1-\delta}}\frac{N_{\rm psr}(E)}{N_{\rm tot}(E)},
\end{equation}
with $N_{\rm psr}$ and $N_{\rm tot}$ the pulsar and total $e^\pm$ spectra. The dipolar anisotropy $\Delta$ is defined as
\begin{equation}\label{eq:aniso}
\Delta = \frac{N_f - N_b}{N_f + N_b}
\end{equation}
where $N_f$ and $N_b$ are the total number of $e^\pm$ observed during a selected ensemble of observations pointing within the sky hemisphere centered on the pulsar ($N_f$) and during a second ensemble of observations {\em with the same collective effective exposure as the first ensemble}, pointing within the opposite hemisphere ($N_b$).

It is worth noting that this calculation of the anisotropy from a single pulsar is overly simplistic, as ignores several possible complicating effects. For instance, the corresponding anisotropy might be washed out by effects such as a local magnetic field bubble, the pulsar's proper motion during the age of $e^\pm$ injection, or significant deviations from the simple diffusive propagation setup employed to theoretically estimate the anisotropy \citep{profumo_occams_razor}. On the other hand, anisotropies in the charged cosmic-ray spectrum can also be induced via diffusion in the interstellar medium, for instance by local magnetic field anisotropies \citep{2008APh....29..420D, 2012PhRvL.109g1101G}. While this may produce a spurious detection of an electron/positron anisotropy not due to a nearby primary source, the two effects may be in principle disentangled in the following ways. First, any anisotropy induced by anisotropic diffusion should affect protons and electrons similarly, leading to a strong correlation between observed anisotropies for both species. In the case of a nearby $e^+e^-$ source, which would not produce many protons due to the strong constraints on primary anti-proton production, the morphology of the anisotropy would not be seen in relativistic protons. Second, any anisotropy stemming from particle diffusion is likely to have an anisotropy which depends on the scale of the magnetic field inhomogenities, while the electron anisotropy from a nearby source will have an energy dependent anisotropy which scales with the positron fraction due to that source. In particular, the anisotropy should disappear above any cutoff energy the primary positron source would possess. Lastly, inhomogenities in diffusion parameters are likely to appear as hotspots~\citep{2008APh....29..420D, 2012PhRvL.109g1101G} or streams \citep{2012arXiv1210.8180K} in the data, an anisotropic signature from which is distinct from the dipole dominated term stemming from nearby sources.

We now turn to the question of how to search for an anisotropy in the cosmic-ray $e^\pm$ flux with ACTs. The most significant uncertainty in the determination of the cosmic-ray $e^\pm$ spectrum with ACTs is the efficiency of cosmic-ray proton rejection. This is the dominant systematic error because the flux of cosmic-ray hadrons dominates the lepton flux by several orders of magnitude.  While observations of $\gamma$-ray point sources are able to employ the isotropy of the cosmic-ray signal in order to control this background, measurements of the cosmic-ray $e^\pm$ flux must instead determine the hadronic or electromagnetic nature of each individual observed shower. To this end, the H.E.S.S. collaboration has adopted a random forest approach \citep{random_forest, bock_zeta_parameter} intended to convert information about the observed shower into a parameter $\zeta$ which describes the extent to which the shower is \emph{electron-like}.  The parameter $\zeta$ is determined in the range of 0~--~1, with larger numbers indicating a better fit to Monte Carlo models of electron showers. While the $\zeta$ parameter is highly energy-dependent, in many situations its discriminating power is significant enough to produce an electron population which dominates the hadronic background at high $\zeta$ values. We note that even for  moderate values of $\zeta$, the contribution from heavier nuclei is entirely negligible.

While a proper selection of $\zeta$ is important so that the cosmic-ray $e^\pm$ population produces a reasonable portion of the total cosmic-ray signal, searches for anisotropy are significantly less affected by errors in hadron rejection compared to measurements of the total $e^\pm$ spectrum \citep{hess_high_energy, hess_low_energy}. Assuming that both the misidentified cosmic-ray proton and background cosmic-ray $e^\pm$ fluxes are isotropic, the fraction of the background which stems from each is irrelevant in searches for $e^\pm$ anisotropies. Instead, the measurable quantity is the fraction of the total cosmic-ray flux (with $\zeta$~$>$~0.9) which stems from an anisotropic candidate pulsar. We can calculate the total contribution to the detected cosmic-ray flux with an ACT as:
\begin{equation}
N_{tot} = (N_{psr} + N_\gamma) + (N_{e, iso} + N_{p}),
\end{equation}
where $N_{psr}$ is the number of cosmic-ray leptons produced by the pulsar, $N_\gamma$ is the number of $\gamma$-rays observed as electromagnetic showers in the instrument, $N_{e, iso}$ is the number of $e^\pm$ from sources which are highly isotropized before diffusing to the solar position, and $N_p$ is the number of protons misidentified as electromagnetic showers. We note that sources in the first parenthesis are measurably anisotropic, while the second set of sources are assumed to be perfectly isotropic in what follows. In our formal calculation, we will ignore the parameter $N_\gamma$, noting that the $\gamma$-ray flux is highly subdominant to the $e^\pm$ and proton fluxes. However, it is important to note that some $\gamma$-ray components, such as the Galactic diffuse emission, are a potentially important source of systematic errors, because their anisotropy is of order unity. We discuss this systematic effect and the impact of our assumption at length in Section~\ref{sec:diffuse}.

If $N_\gamma$~$\sim$~0, the dipole anisotropy can be written in terms of the pulsar contribution to the total cosmic-ray flux as:
\begin{equation}
\Delta = \frac{N_f - N_b}{N_f + N_b} = \frac{N_{psr, f} - N_{psr, b}}{ N_{psr, f} + N_{psr, b} + 2(N_{e, iso} + N_{p})},
\end{equation}
where, consistently with the notation introduced above in Eq.~(\ref{eq:aniso}), $N_{psr, f}$ and $N_{psr, b}$ are the total number of $e^\pm$ from the candidate pulsar from the two ensembles of observations when the telescope is pointing at locations in the hemisphere oriented directly towards, or directly away from the pulsar (note that the factor 2 in the denominator stems from our assumption of identical effective exposures for the two ensembles of observations). An anisotropy measurement is then significant at the 2$\sigma$ level when:
\begin{equation}
\Delta > 2 \frac{\sqrt{N_{avg}}}{N_{avg}},
\end{equation}
where $N_{avg}=(N_f + N_b)/2$ is the average number of showers observed during each ensemble of observations. Using this model, we can now estimate the anisotropy level, as a function of energy, which can be detected with an ACT, and compare it with the current observational constraints from Fermi-LAT and AMS-02. 


The key input to our estimate of the sensitivity to an $e^\pm$ anisotropy is the determination of $N_{avg}$ which, in turn, depends on the $e^\pm$ flux measured by the ACTs. We start with predictions on the limits which can be set using archival data from the H.E.S.S. telescope. We note that other ACTs, such as VERITAS~\citep{veritas}, can in principle make very similar measurements. However, details of VERITAS measurements of the electron spectrum are not publicly available, and we therefore employ H.E.S.S. for our case study.

H.E.S.S. has presented two different results on the cosmic-ray $e^\pm$ spectrum. The first result concerns the high-energy spectrum, in the range of 1--4~TeV~\citep{hess_high_energy}. In this regime, H.E.S.S. reports an effective area of approximately 5$\times$10$^4$~m$^2$. The study employed 239h of live-time, producing an effective exposure of 8.5$\times$10$^7$~m$^2$~sr~s. In this study, H.E.S.S. found a best-fitting $e^\pm$ spectrum 
\begin{equation}
{\rm d}N/{\rm d}E = k(E / 1\ {\rm TeV})^{-\Gamma},
\end{equation}
with k = 1.17$\pm$0.02~$\times$~10$^{-4}$~TeV$^{-1}$m$^{-2}$sr$^{-1}$s$^{-1}$, and $\Gamma$= 3.9$\pm$0.1 (stat). Given the H.E.S.S. effective exposure, this implies the detection of 3370 $e^\pm$. Indeed, using the Dexter package\footnote{http://dexter.sourceforge.net/}, we find that H.E.S.S. observed 3100 $e^\pm$ with $\zeta$~$>$~0.6. Of these events, 2600 were found to have $\zeta$~$>$~0.9, giving this cut a relative acceptance of 0.77. Interestingly, while H.E.S.S. reports the detection of 7610 protons with $\zeta$~$>$~0.6, only 2470 of these protons have $\zeta$~$>$~0.9, yielding a sample which is, therefore, electron dominated.

Additionally, H.E.S.S. investigated the low energy $e^\pm$ spectrum in \citet{hess_low_energy}, in order to investigate the spectral bump observed by ATIC~\citep{atic_electron_bump}. In this study, H.E.S.S. employed only 77 hours of livetime, with a calculated effective area of 3$\times$10$^4$~m$^2$, for a total effective exposure of 2.2$\times$10$^7$~m$^2$sr~s. H.E.S.S. obtained a best-fitting broken power law spectrum:
\begin{equation}
{\rm d}N/{\rm d}E = k(E/E_b)^{-\Gamma_1}(1 + (E/E_b)^{1/\alpha})^{-(\Gamma_2-\Gamma_1)\alpha},
\end{equation}
with a best fit values $E_b$=0.9~$\pm$~0.1~TeV, $k$~=~(1.5~$\pm$~0.1)~$\times$~10$^{-4}$~TeV$^{-1}$m$^{-2}$sr$^{-1}$s$^{-1}$, $\Gamma_1$=3.0$\pm$0.1 and $\Gamma_2$=4.1$\pm$0.3. This indicates that H.E.S.S. observed approximately 7660 $e^\pm$ with energies between 340--700 GeV, the range in which they provide the observed distribution over $\zeta$. Unfortunately, in this low energy regime, the $\zeta$ parameter is less efficient at differentiating the cosmic-ray electron and proton spectrum, and H.E.S.S. observes only 2900 $e^\pm$ with $\zeta$~$>$~0.9 compared to 5020 protons. 

\section{Results}
\label{sec:results}

A remarkable aspect of ACT searches for $e^\pm$ anisotropy is that no dedicated observations are required in addition to archival data. As long as the target pulsars fall in regions of the sky such that, on average, the ACT spends a significant fraction of observing time in hemispheres containing and not containing the pulsars, a sizable fraction of the total archival observation time can be used to search for $e^\pm$ anisotropies. The two pulsars we consider here (Monogem is at (l,b) = (201.1, +8.3), and Geminga is at (l,b) = (195.1, +4.3)) fall, for example, in a favorable part of the sky for H.E.S.S. observations\footnote{We note that both pulsars are located relatively close to the Crab nebula, which is closely monitored by all ACTs.}. In practice, the available observation time is the smallest of the collective time the telescope spent observing in hemispheres containing and non-containing the relevant candidate pulsar. 

In order to predict the H.E.S.S. constraints which could be set using current data, we need to make a multitude of assumptions. We caution the Reader that several relevant figures have been approximated quite roughly, as an accurate calculation is impossible without further information about the H.E.S.S. telescope as well as its pointing history.  We first approximate the total available livetime of the instrument to be approximately 3000h for searches off of the Galactic plane (where the $\gamma$-ray background is much more significant), noting that the recently published H.E.S.S. line search included 1153h of livetime taken during studies of extragalactic objects between 2004-2007~\citep{hess_hours_spent_on_extragalactic_observation}, and that additional surveys (such as those of dwarf spheroidal galaxies) will also provide excellent targets for $e^\pm$ anisotropy searches. In order to calculate the effective area of the H.E.S.S. instrument, we take the model of \citet[Fig 2a.]{hess_effective_areas}, assuming an average zenith angle of 45$^\circ$, and we take the H.E.S.S. field of view to cover 3.8$\times$10$^{-3}$ sr. We additionally consider a future global observation time of 5000h, including new data taken by the H.E.S.S.-II telescope. 

We note that this combination provides an effective exposure at 340~GeV which exceeds that reported by \citet{hess_low_energy} by approximately a factor of 8, presumably due to additional cuts regarding the nature of the electron shower and the removal of $\gamma$-ray point source contamination. We note that this mismatch is slightly worse when comparing the 1~TeV calculation with that of \citep{hess_high_energy}, as additional cuts are made in this study. In what follows we degrade our calculated effective exposure by an \emph{ad hoc} factor of 5. We note that this may be overly conservative, as cuts optimized for this study could allow for a much larger effective area, since this analysis is by its nature less susceptible to systematic errors regarding proton contamination.

\begin{figure}
                \plotone{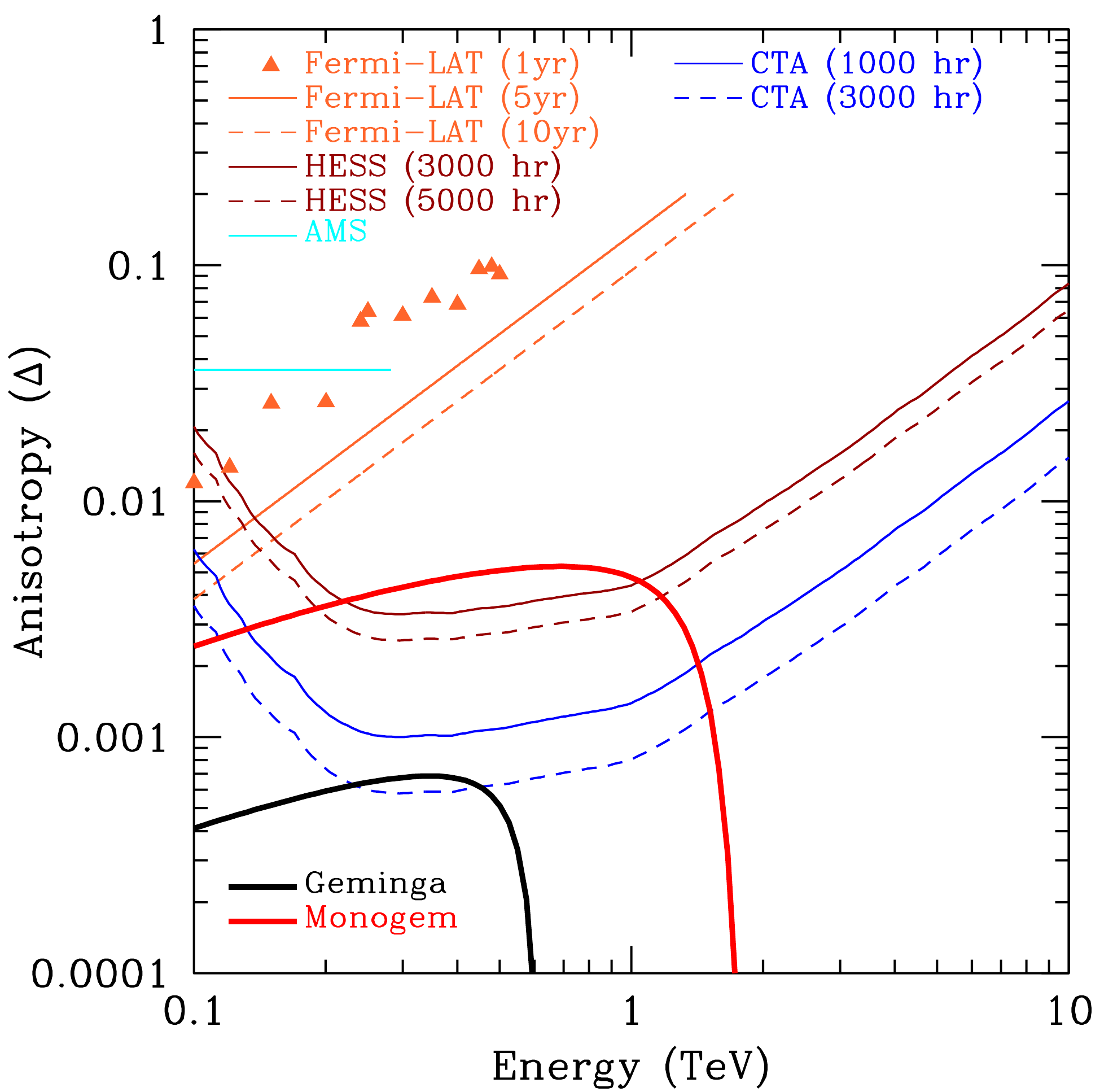}
                	\caption{ \label{fig:limits} The limits on the cosmic-ray $e^\pm$ anisotropy one year of Fermi-LAT data (orange triangles) and those recently reported by AMS (cyan), as well as the predicted limits from 5 and 10 years of Fermi-LAT observations (orange solid and orange dashed), along with the predicted limits from 3000 and 5000 hr of H.E.S.S. observations (maroon solid and maroon dashed), as well as predicted limits from 1000 and 3000 hours of CTA observations (blue solid and blue dashed). These limits are compared with the predicted fluxes for models of the Geminga (black solid) and Monogem (red solid) pulsars which correctly explain the positron excess observed by AMS-02. We note that limits from the Fermi-LAT are technically set based on a minimum energy E, rather than a traditional E dN/dE, a difference which is less important given the steeply falling $e^\pm$ flux.}
\end{figure}

In order to calculate the effect of the $\zeta$ parameter on the separation of electrons and protons, we make a cut of $\zeta$~$>$~0.9, and use the values given by \citep{hess_high_energy} and \cite{hess_low_energy} to calculate the loss of effective area after this cut is applied at energies of 1~TeV and 340~GeV respectively, which correspond to the low end of the energy range employed in each analysis. Thus, we assume a relative electron acceptance of 0.38 at energies below 340~GeV and 0.77 at energies above 1~TeV, linearly interpolating between these values at intermediate energies. Finally, using the observed proton background at $\zeta$~$>$~0.9 for each observation, we assume an irreducible proton background which is 1.73 (0.95) times larger than the total $e^\pm$ flux at energies of 340 GeV (1 TeV), and again linearly interpolate between these values. 

In order to compare our results with the projected limits from five years and ten years of Fermi-LAT data, we take the limits from one year of data obtained by \citet{fermi_1yr_electron_anisotropy} and calculate a best fit power law to the 95\% upper limits of $\Delta$~$<$~0.30~(E / 1 TeV)$^{1.39}$ and then reduce the limits by the square-root of the additional exposure time, in order to  extrapolate optimistic results where the exclusion limits are dominated by statistics only. We note that this result may slightly improve, in light of new data-taking algorithms such as Pass-8, which are likely to increase the effective area to electron showers~\citep{pass8}. We additionally compare our result with the 95\% confidence limit of $\Delta$~$<$~0.036 measured by AMS-02, though we note that no energy scaling was provided with this value.

In addition to these H.E.S.S. and Fermi-LAT observations, we predict observations from the upcoming Cherenkov Telescope Array (CTA). While the parameters of CTA are unknown, for the purposes of repeatability we assume the following educated guesses for the relevant parameters. We take the effective area to be an order of magnitude greater than the H.E.S.S. telescope, and the field of view to be a factor of (3/2)$^2$ larger, corresponding to the models suggested in~\citet{ctaspecs}. Additionally, we note that the multiple telescopes of CTA greatly enhance its hadronic rejection capabilities, thus we add an \emph{ad hoc} factor of two decrease in the relative hadronic flux for CTA observations. All other parameters (including the $\zeta$ cut and the factor of 5 degradation in the effective exposure seen in H.E.S.S. observations) are left the same. 

In Figure~\ref{fig:limits} we show the current limits given by 1 year of Fermi-LAT data, the recently released AMS-02 limits on the $e^\pm$ anisotropy, as well as the projected limits from 3000h and 5000h of H.E.S.S. observations and from 1000h and 3000h of CTA observations, compared to the projected anisotropies of the Geminga and Monogem pulsars with the same setup as described in Section \ref{sec:flux} and shown in Fig.~\ref{fig:flux}.  Note that the observation times are meant here as total duration of the two ensembles of observations in the hemispheres towards and opposite the direction of the pulsar of interest. For example, for 3000 hr, in the notation introduced in the previous section, the global duration of the ensemble of observations yielding $N_{f,b}$ is 1500 hr each. For all models we assume a total $e^\pm$ flux given by the best fitting power-law given by \citet{hess_low_energy}. We find that current H.E.S.S. archival observations have the potential of observing the anisotropy induced by Monogem, while CTA observations will be necessary in order to observe any anisotropy from Geminga. The predicted level of anisotropy from either pulsar under consideration here is fully compatible with the limits from Fermi-LKAT and from AMS-02.





\section{The Diffuse Gamma-Ray Background}
\label{sec:diffuse}

A serious concern in the identification of the $e^\pm$ anisotropy stems from the misidentification of the $\gamma$-ray background. Unlike hadronic showers, $\gamma$-ray and electron showers are nearly identical in nature, with the only observable difference pertaining to a slightly different value of the reconstructed shower maximum (X$_{max}$). However, this parameter is very difficult to accurately measure with ACTs, and thus only approximately 20\% of the $\gamma$-ray signal can be effectively eliminated before significantly cutting into the electron effective area~\citep{hess_high_energy}.

There are two important sources of $\gamma$-ray contamination that we need to consider here. The first corresponds to the extragalactic $\gamma$-ray background (EGRB), which is the the brightest component of the very high energy $\gamma$-ray sky, and is thus the major contributor to the uncertainty in the $e^\pm$ spectrum. However, the EGRB is extremely isotropic, especially on large scales, and does not contribute to the $e^\pm$ dipole anisotropy. The EGRB can thus be treated as a statistical background, as in the case of the hadronic contribution. In order to calculate the size of this effect, we extrapolate the calculated EGRB from Ferm-LAT observations, which indicate the EGRB to follow a power law 
\begin{equation}
{\rm d}N/{\rm d}E = k(E/1 TeV)^{-\gamma},
\end{equation}
with $k=3.3 \times 10^{-7}$ TeV$^{-1}$m$^{-2}$s$^{-1}$sr$^{-1}$ and $\gamma$~=~2.41~\citep{fermi_egb}. This implies that at energies near 1~TeV, the EGRB is subdominant to the $e^\pm$ contribution by a factor of $\sim$250. Thus, we do not consider the effects of the EGRB on our results. 

A larger concern stems from Galactic diffuse $\gamma$-ray emission, which (due to our position in the Galaxy) has a dipole anisotropy of order unity. In order to ascertain the influence of this component on our results, we employ \emph{Galprop} models of the Galactic diffuse emission, following the same parameters employed previously in the text. At high energies, we find the dominant source of diffuse $\gamma$-ray emission to stem from $\pi^0$ decays produced when energetic protons collide with Galactic gas. Eliminating observations within 5$^\circ$ of the Galactic plane, as well as a 10$^\circ \times$10$^\circ$ box around the Galactic center, we find a maximum $\gamma$-ray flux of 1.33 $\times$ 10$^{-6}$ TeV$^{-1}$m$^{-2}$s$^{-1}$sr$^{-1}$, which falls below the total lepton flux by a factor of 60. Additionally, by setting somewhat stronger constraints to limit the regions where Galactic gas may produce bright emission (setting $|b|$~$>$~10$^\circ$ and eliminating a box of 20$^\circ$ around the Galactic center), we find a maximum $\gamma$-ray flux of 9.0 $\times$ 10$^{-6}$ TeV$^{-1}$m$^{-2}$s$^{-1}$sr$^{-1}$, which is subdominant by more than two orders of magnitude.

Since the diffuse $\gamma$-ray flux contributes with order 1 anisotropies, its contribution to the overall anisotropy can still exceed the predicted $e^\pm$ anisotropies, which are likely to be less than 1\%. However, we note that this systematic error can be controlled in two independent ways. First, the dipole anisotropy stemming from $\gamma$-ray emission will produce a peak in emission pointing towards the Galactic center, while the Monogem and Geminga pulsars are both located in the Galactic anti-center region. While the pulsars may have moved considerably since their $e^\pm$ flux was produced \citep{profumo_occams_razor}, it is unlikely that either pulsar was located in a position aligned with the Galactic center. Secondly, we note that the morphology of $\pi^0$ decay is relatively energy independent, and is extremely well mapped at energies below 100~GeV by the Fermi-LAT, thus a statistical comparison of any anisotropy with the Fermi-LAT diffuse $\gamma$-ray sky should immediately indicate, or rule out, a $\gamma$-ray origin. This method can additionally be used to remove portions of the $e^\pm$ anisotropy which correlate strongly with locations of high diffuse $\gamma$-ray luminosity. 

\section{Systematic Uncertainties}

There are several systematic issues which further complicate the analysis presented here, and which will need to be carefully considered in any experimental effort to detect an electron anisotropy. First, unlike the Fermi-LAT, which achieves nearly uniform sky coverage, ACTs observe non-uniform patches of the sky. While no direct observations of candidate pulsars are necessary in order to perform the calculation suggested here, the ability for ACTs to determine the $e^\pm$ anisotropy depends sensitively on the effective exposure of observations in the direction of each pulsar. For Geminga and Monogem, we note that their relative proximity (within 10$^\circ$ on the sky), and their location near the galactic anticenter, makes this observation reasonable. 

Additionally, we note that our assumption of proton isotropy does not hold for the very precise measurements ACTs are capable of making. While the exact magnitude of this anisotropy is difficult to ascertain, as measurements by MILAGRO \citep{2009ApJ...698.2121A} and ICECUBE \citep{2009arXiv0907.0498A, 2010ApJ...718L.194A} measured the proton anisotropy for median proton energies of 6~TeV and 20~TeV respectively. However, both experiments observe anisotropies on the order of 0.1\%, of similar magnitude to the predicted lepton anisotropy. While a separation of these possible anisotropies requires careful experimental consideration, some approaches exist which may help separate these components. For example, we note that the proton measurement should be exquisitely measurable by ACTs, with almost no electron background contamination, simply by evaluating showers in ranges of $\zeta$ where the proton flux dominates by orders of magnitude. The resulting proton anisotropy can then be subtracted from the anisotropy of the electron measurement by employing a template fit. We note that the pulsar sources which we expect to dominate the electron anisotropy emit a negligible proton population, making the measured anisotropies largely uncorrelated. We also note that the proton anisotropy is observed to exist in localized hotspots, which may be separable from the dipole dominated electron anisotropy our models suggest.

Finally, we note that anisotropies can be induced due to instrumental systematics, such as uncertainties in the effective acceptance could artificially induce an instrumental sensitivity. We note that while overall errors in the calculated effective area are immaterial to calculations of the anisotropy, both optical degradations (especially in the photo-multiplier tubes) and night-to-night optical variations (based on the the background starlight and ambient temperature) can have a large effect on the measured anisotropy. Again, controlling these systematics requires careful experimental work in evaluating the observed electron spectrum under variations of many known parameters, and we can only suggest mechanisms which may make this measurement possible. We note that the event-shuffling technique employed by the Fermi-LAT collaboration to study anisotropies can, in principle also be employed with ACT observations in order to test for systematic biases~\citep{fermi_1yr_electron_anisotropy}. Furthermore, the large statistics in the proton anisotropy detected above can be used to independently test the systematic uncertainties in the instrumental effective area.

\section{Discussion and Conclusions}
\label{sec:discussions}

We showed that the recent measurement of the positron fraction by AMS-02 is in excellent agreement with the simplest models of a diffuse cosmic-ray background with contributions from a single, nearby pulsar emitting $e^\pm$ pairs with a power law spectrum with spectral index close to 2. Since this stands as the minimal explanation which fits all current data, arguments along Occam's Razor would make it a preferable model. Secondly, we showed that this scenario is differentiable from both models of dark matter annihilation, as well as from models invoking a multitude of pulsars, based on the induced anisotropy in the cosmic-ray $e^\pm$ arrival directions that would stem from one  single (or very few) bright, nearby source. Finally, we have proposed a novel method to observe the predicted $e^\pm$ anisotropy, using the wide collecting area of ACTs and large ensembles of archival observations in order to collect a substantial population of cosmic-ray $e^\pm$ from all directions in the sky, and argued that this method is capable of setting the strongest constraints on the cosmic-ray $e^\pm$ anisotropy. 

We note that, under certain assumptions on the dark matter mass and annihilation final states, relatively large anisotropies can be produced even in the context of dark matter annihilation as the source of the anomalous positron fraction \citep{2010arXiv1012.0041B}. However, we note two possible complications for dark matter models containing significant anisotropies. First, the predictions of Ref.~\citet{2010arXiv1012.0041B} are significantly affected by AMS-02 results, since very special dark matter final states and masses are compatible with the high-statistics AMS-02 data (see e.g. Ref.~\citet{2013arXiv1304.1840C}); it is unclear whether the large levels of anisotropy found in \citet{2010arXiv1012.0041B} are still viable after the AMS-02 results. Second, the largest anisotropies from dark matter stem from a highly clumpy dark matter density distribution, as discussed in Ref.~\citet{2010arXiv1012.0041B}. However, a bright local clump would have been observed in gamma rays, as pointed out e.g. Ref.~\cite{Brun:2009aj}; it is thus unclear whether any large anisotropy would be actually compatible with gamma-ray observations.

Finally, while this work has focused on prospects for H.E.S.S., the position of VERITAS in the northern hemisphere makes it an equally good telescope with which to measure the $e^\pm$ anisotropy independently of the H.E.S.S. results. While systematic effects may greatly degrade the sensitivity of each instrument towards the detection of electron anisotropies, we note that  ACTs are predicted to set constraints on the TeV scale electron anisotropy which are nearly two orders of magnitude stronger than the predicted 10-year constraints from the Fermi-LAT. This means that even if systematic issues greatly degrade the limits which can be set by ACT instruments, they are still likely to probe interesting regions of the parameter space.

\acknowledgements
We would like to thank Amy Furniss for valuable discussions on the capabilities of ACT telescopes. This work is partly supported by  the Department of Energy under contract DE-FG02-04ER41286 \newline
      
\bibliography{act}

\end{document}